\documentclass[11pt,twoside]{article}
\usepackage{asp2010}
\resetcounters

\begin{document}

\title{Structure and Dynamics of the Milky Way: The Evolving Picture}
\author{T. Foster$^1$ and B. Cooper$^1$
\affil{$^1$Dept. of Physics \& Astronomy, Brandon University, 270-18 St. Brandon, Manitoba R7A6A9 CANADA}}

\begin{abstract}
From the inception of radio astronomy, the study of the Interstellar Medium has 
been both aided and frustrated by one fact: we are right within it! Our 
embedded perspective is favourable to observing kinematic and structural 
details that are likely to remain inaccessible in external galaxies for some 
time, but not so to gaining a birds-eye view of our Galaxy's structure and 
motions on the largest scales. The Canadian Galactic Plane Survey (CGPS) is the 
original survey with the ability to image both details and the 
{}``big picture''. We will broadly review what is known of the Milky Way, and 
focus on large-scale ISM structure and dynamics that the CGPS and immediate 
family of surveys depicts particularly well (e.g. spiral structure, the 
rotation curve, density waves, rolling motions, the warp 
\& scalloping). We also highlight areas where puzzles still exist (e.g. outer 
spiral structure, the question of shocks and rolling motions in 
the Milky Way), and offer some new insights (e.g. multiple shocks in the
2$^{\textrm{nd}}$ quadrant; a radially varying spiral pattern 
speed in the disc) that demonstrate what is possible with current and future 
high-resolution 21~cm surveys.
\end{abstract}

\section{Introduction and a Brief History}
Although the meeting that generated these proceedings coincides with the end of 
observing for the Canadian Galactic Plane Survey \citep{tayl03} and 
celebrates the completion of radio astronomy's original "small-structure: 
big-picture" survey, it is more properly viewed as a rest-stop along the 
journey that all ISM astronomers share in. Discoveries made at the survey's end 
will undoubtedly be as profound as the ones we have made along the way and will 
continue to make. The journey for those who have been involved with the CGPS 
has also been marked with images of an extraordinary beauty only a well-sampled 
$u,v$ plane could provide. This review will take stock of discoveries we have 
accumulated up to this point and synthesize a contemporary picture of the Milky 
Way's large-scale structure and dynamics. It follows a talk on the same, online 
at www.novell.brandonu.ca/$\sim$fostert/foster.ppt. As the author is a progeny 
of the CGPS, the review is naturally focussed on aspects of the Milky Way 
(hereafter MW) that are depicted well in it and its siblings VLA-GPS and 
Southern-GPS, which came together in 2001 to form the International Galactic 
Plane Survey, or IGPS.

One theme that the reader may expect to encounter in pursuing or reviewing 
Galactic ISM structure and dynamics studies is an often frustrating 
interdependency between them. The cause is in the brightness temperature unit 
$T_B(v)dv$ (K), which descends from both line-of-sight (hereafter LOS) density 
variations $n_{HI}(r)$ and inflections in the velocity field $dv/dr$. This 
{}``degeneracy'' frustrated the 1970's pioneers of the field; for 
example, \citet{burt71} attempted to say something about the distribution of 
Galactic hydrogen by first assuming its motion (rotation only), whereas Rohlfs' 
\citeyear{rohl74} investigation of the velocity field could only proceed by 
assuming some density structure for the disk. Along with the confusion from the 
overwhelming amount of emission, the nearly indistinguishable observational 
signatures of different density wave theories, and the lack of 
computing power at the time, this led to early efforts to model spiral 
structure with \ion{H}{i} being largely abandoned by the late 1970's. The 
demise was foreshadowed by Frank Kerr's comments on Galactic \ion{H}{i} spiral
structure at the 1970 {}``Spiral Workshop'' \citep[an account of which is 
by][]{simo70}, where he introduced 21~cm emission as "the way to study the 
whole Galaxy", only to concede later that "there is so much 21~cm emission that 
we're confused by it" \citep[see][]{lock02}. As you will see, the picture 
remains elusive, and to this day there are doubters that \ion{H}{i} is at all a 
reliable tracer of spiral structure. 

\section{Fundamental Constants}
\label{fundconsts}
The fundamental equation of Galactic structure relates the observed Doppler 
velocity along the line-of-sight $v_{LOS}$ to the Galactic orbital motion 
$\Omega$ of an ISM element:
\begin{equation}
\label{fundeqn}
v_{LOS}=R_0\left[\Omega\left(R \right) - \Omega_0\right]\textrm{sin}\ell~\textrm{cos}b
\end{equation}
where the fundamental structural constant $R_0$ is the Solar Galactocentric 
radius and the fundamental dynamical constant $\Omega_0=v_0\textrm{/}R_0$ is 
the angular velocity of the Sun (with respect to the LSR). Mapping out 21~cm 
Galactic structure and dynamics requires each of $\Omega$ and $R$ to be 
measured independently for the element. Since $v_{LOS}$ is measured, $\Omega$ 
is an observable quantity independent of $R$, provided both $\Omega_0$ and 
$R_0$ are known. Only if perturbations in $R$ and $z$ are negligible is 
$v_{LOS}$ then a direct observation of the true \textit{rotational} frequency 
$\Omega$ of the element with respect to the Sun and Eqn. \ref{fundeqn} is 
independent of $R$. Unfortunately, these ideal conditions do not often happen 
in the plane. 

\begin{figure}[ht]
\centerline{
\includegraphics[scale=0.28,clip=]{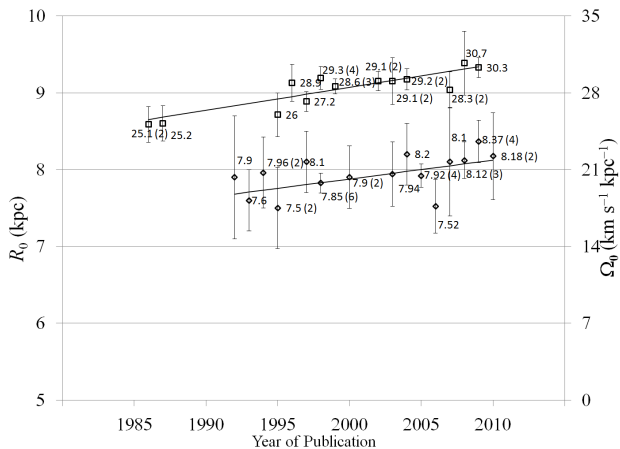}}
\caption{Published Galactic constants $R_0$ ($\diamond$; left 
axis) and $\Omega_0$ ($\square$; right) over time.}
\label{fund_const}
\end{figure}

The angular velocity in the Solar neighborhood $\Omega_0$ is most often 
measured by observing kinematics of stellar tracers such as OB stars, 
Cepheids, open star clusters, or old red giants. All methods either rely on 
observables like the Oort constants, ($\Omega_0=A-B$) or require the distance 
to each tracer (and hence rely on the constant $R_0$). 
It appears that differences in measured values for both constants are subject 
to the observer's bias more so than to random differences from the techniques 
used to measure them. Hence neither $\Omega_0$ or $R_0$ have converged towards 
single values with time. A gradual increase in both is shown in Figure 
\ref{fund_const}, which plots 32 published measurements of $R_0$ and 23 for 
$\Omega_0$. The data (Table 1, available online at 
www.novell.brandonu.ca/$\sim$fostert/table1.pdf) are grouped together 
into one-year bins by calculating the uncertainty-weighted mean for each year 
(labelled in figure). The uncertainty for each bin is the mean published 
uncertainty that year divided by the square root of the number of measurements 
that year (in brackets on figure). The trend in $\Omega_0$ (Pearson coefficient 
r=0.87) is significant and likely attributable in part to the {}``bandwagon'' 
effect where new values will tend to agree with previously published ones 
regardless of the method used to obtain them. A correlation of $R_0$ with 
publication year is also observed (r=0.59; 95\% level). This suggests that the 
same effect in $R_0$ \citep[previously noted by][]{reid93} has carried forward 
beyond 1993, although all data bins in Fig. \ref{fund_const} are within error 
of one another. An error-weighted mean for each constant $R_0=$8.0$\pm$0.4~kpc 
and $\Omega_0=$28.7$\pm$1.1~km~s$^{-1}$~kpc$^{-1}$ is calculated, though it is 
likely that neither sample is complete.
For this review we adopt the modern constants of \citet{reid09} based on 
trigonometric parallax observations of star formation regions: 
$R_0=$8.4$\pm$0.6~kpc and $\Omega_0=$30.3$\pm$0.9~km~s$^{-1}$~kpc$^{-1}$ (or a 
Solar linear velocity of $v_0=$254~km~s$^{-1}$).

\section{Galactic Dynamics}
\subsection{Milky Way Circular Rotation}
The largest scale dynamics in the Milky Way are those that reflect the response 
of the matter to the overall gravitational potential $V_0\left(R\right)$ 
created by the mass distribution (stars plus ISM). For circular rotation 
the potential is related to observables $\Omega$ and $R$ by 
$dV_0\textrm{/}dR=\Omega^2R$. Circular orbits in galaxies are found with 
$\Omega\propto~R^{-1}$ or a constant circular speed, and never at the extremes 
$\Omega\propto R^{-3/2}$ (Keplerian) and $\Omega=const$ (solid-body rotation). 
The actual shape of MW rotation near the Sun can be assessed with the Oort 
constants, or equivalently the frequency of radial oscillations about the 
circular orbit's guiding centre $\kappa_0=\sqrt{-4B(A-B)}$. The ratio of 
epicyclic-to-orbital frequency, $\kappa\textrm{/}\Omega$ is $\sqrt{2}$ for flat 
circular speed curves. This is in fact what is observed near the Sun: for 
example using open star clusters \citet{lepi08} find 
$\kappa_0=$43$\pm$5~km~s$^{-1}$~kpc$^{-1}$. Then 
$43\textrm{/}30.3\sim\sqrt{2}$, suggesting MW rotation is quite flat in the 
vicinity $R=R_0$.

\subsubsection{Inner Galaxy Rotation}\label{igrtext}
\begin{figure}[ht]
\centerline{
\includegraphics[scale=0.19,clip=]{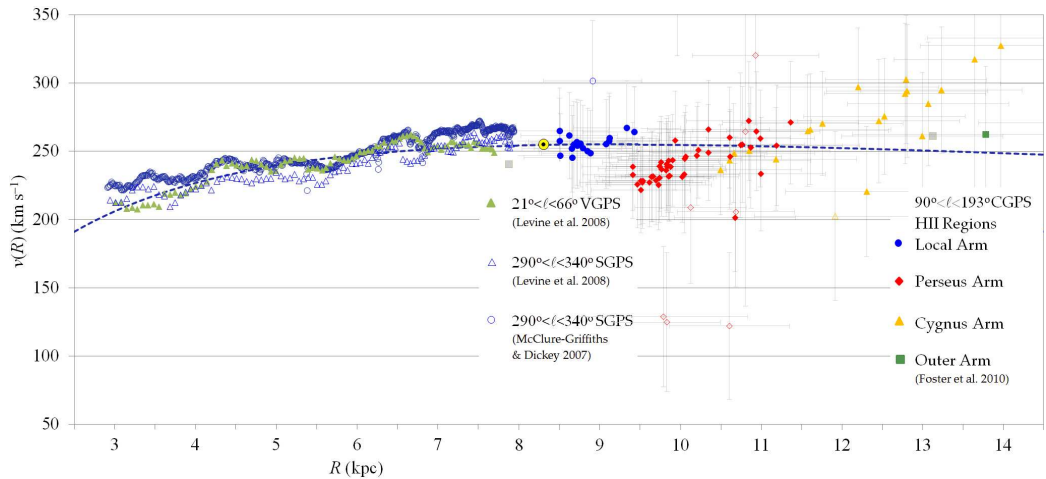}}
\caption{MW rotation curve between $R=$3-14~kpc. Inner Galaxy data are 
from the VGPS, SGPS surveys (\ion{H}{i} terminal velocity method) and the Outer 
Galaxy from the CGPS (\ion{H}{ii} regions). The two large grey squares are points from \citet{xue08}.
}
\label{igpsrc}
\end{figure}

Because of extinction and the relatively low velocity dispersion of the
interstellar gas ($\sigma_v\sim$8~km~s$^{-1}$), ubiquitous radio ISM gases
(\ion{H}{i}, $^{12}$CO) are better tracers of MW rotation in the plane than are 
stars, which are veritable bullets with $\sigma_v\sim$30~km~s$^{-1}$. Inside of 
the Solar circle, the rotation curve is directly observable by isolating the 
part of the spectral profile emitted by gas at the \textit{tangent point}, 
where the LOS passes closest to the Galactic centre. 
At this point $R=R_0\textrm{sin}\ell$, and the observable quantity in Eqn. 
\ref{fundeqn} becomes the circular velocity at this radius 
$v(R_0\textrm{sin}\ell)=v_{t}+v_0\textrm{sin}\ell$.
Success at isolating emission from gas there depends on 1) the method used to 
locate $v_t$ within a profile, and 2) the ability to distinguish it from the
innumerable clouds, expanding shells, continuum and self-absorption features 
and other small-scale features that display their own kinematics. The best 
methods model the profile near $v_t$ with some empirical or 
physically-motivated function \citep[e.g.][]{rohl87} rather than simply placing 
$v_t$ at some repeatable threshold within the profile \citep[the $T_B$ 
half-maximum;][for example]{luna06}. As for 2), only high-resolution data like 
those produced by the IGPS can help. The two best \ion{H}{i} rotation 
curves for $R<R_0$ on the market today are extracted from VGPS 
(1$^{\footnotesize{\textrm{st}}}$ quadrant, or QI) and SGPS (QIV) data: 
\citet{levi08,mccl07}. These are compared in Fig. \ref{igpsrc} after rescaling to 
our adopted constants in Sec. \ref{fundconsts}.
The QIV curve of \citet{mccl07} illustrates the benefits of high-resolution 
data: it is densely sampled and makes good use of the high-resolution SGPS data 
by avoiding continuum absorption features. However their curve is some 
$\sim$7~km~s$^{-1}$ above that of \citet{levi08}, a difference that comes from 
modelling the profile with a complementary error function, which restricts 
$v_t$ to be down the wing near the half-maximum in the profile. \citet{levi08} 
use a model profile which is more physically consistent with circular motion at 
the tangent point \citep[that of][]{celn79}, and locates $v_t$ higher up the 
wing at about 90\% of the peak, where it is generally found in synthetic 
observations of hydrodynamics galaxy simulations. They also measure QI with 
VGPS data, and their results show the long-known asymmetry between the QIV and 
QI curves. For their study of the $z$-dependence of $v(R)$ they smear VGPS and 
SGPS data into 0\fdg5 bins, and their curves at $z=$0 thus suffer from higher 
scatter than \citet{mccl07}. The CGPS has yet to be used to measure QI rotation 
for 50\deg$\leq\ell\leq$90\deg ($R=$6.4-8.4~kpc) but should provide an 
outstandingly clean curve if care is taken to avoid \ion{H}{i} continuum- \&
self-absorption.

\subsubsection{Outer Galaxy Rotation}\label{ogrtext}
Outside of the Solar neighborhood, MW rotation in the disc is traceable to 
great distances with the radio ISM, out to the edge of the \ion{H}{i} disc 
($R\sim$20~kpc). Because the term 
$dv/dr\rightarrow~0$ at these distances, the edge of the \ion{H}{i} surface 
density as measured with Eqn. \ref{fundeqn} is very sensitive to the shape of 
the rotation curve \citep{lock02}. In turn, the slope of the far outer curve 
measured with Eqn. \ref{fundeqn} using discrete kinematic tracers depends 
strongly on the adopted values of constants $\Omega_0$ and $R_0$, and on 
uncertainties in $\Omega$ and $R$. In particular, uncertainties in $R$ for 
spiral arm tracers like \ion{H}{ii} regions bias part of the curve to appear to 
rise like solid-body rotation \citep[$\Omega\sim const$][]{binn97}; in other 
words a rising circular speed curve could be observed when it is in fact 
constant or falling through the outer Galaxy. This effect is plainly visible in 
Perseus and Cygnus arm \ion{H}{ii} regions plotted in Figure \ref{igpsrc} 
appearing in each arm as a rising $v(R)\propto R$ trend. This bias is 
undoubtedly responsible for the rise observed in the classic Outer Galaxy 
rotation curve of \citet{bb93} traced with bright \ion{H}{ii} regions. Indeed 
it must be altering the measured shape of all rotation curves constructed 
principally from spiral arm tracers. Generally, contemporary published Outer 
Galaxy curves \citep[e.g.][]{russ07,hou09} are mainly compiled from stellar 
distances and velocities from across the literature, with little new data 
\citep[one exception is][]{russ07}. These curves generally show constant 
circular speed (i.e. $\Omega\propto R^{-1}$) well out to the edge of the Milky 
Way's stellar disc ($\sim$15~kpc). Near-constant circular speed has been 
confirmed by modelled halo potentials from SDSS observations \citep{xue08} out 
to an extraordinary $R=$50~kpc where it declines to only $\sim$200~km~s$^{-1}$.

A feature common to all Outer Galaxy curves based on discrete kinematical 
tracers is a large scatter in $R$ and (to a lesser extent) in $\Omega$, which 
overwhelms any detail in the curve of the kind we see in the Inner Galaxy. To 
see these details, 1-arcminute CGPS \ion{H}{i} and $^{12}$CO line data can be 
a great advantage, measuring the systemic velocity from the thin edges of 
shells that surround the ionized gas and stars. Where both \ion{H}{i} and CO 
shell edges are seen, one gets two measures of $v_{LOS}$ which are 
perpendicular to any motions from the shell's expansion along the LOS. To 
demonstrate this advantage we will refer to measurements of velocities and 
distances of some 100 \ion{H}{ii} regions across much of the CGPS 
(90\deg$\leq\ell\leq$193\deg) \citep[from][for this review we will refer to 
these tracers often]{fost10}. In this sample 33 Perseus spiral arm objects are 
also found in \citet{bb93} (excluding those with $\ell=$170-190\deg due to the 
1/sin$\ell$ projection effect) and in that paper, these show a mean angular 
velocity of $\Omega_{ave}=$24.67$\pm$1.04~km~s$^{-1}$~kpc$^{-1}$ (1-$\sigma$). 
Using CGPS data two velocities each ($^{12}$CO \& \ion{H}{i}) are determined 
for 41 Perseus arm \ion{H}{ii} regions in the same longitude range, resulting 
in $\Omega_{ave}=$24.13$\pm$0.69~km~s$^{-1}$~kpc$^{-1}$. In the CGPS we not 
only find velocities for 8 more objects but the scatter about a common 
$\Omega$ has been reduced by 33\% \footnote{Additionally in \citet{fost10} 
distances to OB stars within the \ion{H}{i} and $^{12}$CO shells surrounding 
each \ion{H}{ii} region have been recalculated using a common MK Luminosity 
class and a common ($B-V$) colour calibration, reducing the scatter in $R$ by 
20\%}.

The Outer Galaxy rotation curve $v(R)$ obtained with all 100 regions is shown 
in Fig. \ref{igpsrc}. Fitted to all data in the figure (except $v_t$ from 2007 
and \ion{H}{ii} regions between $\ell=$170-190\deg, marked with open symbols) 
is Eqn. \ref{rceqn} below, which gives a reasonable approximation to observed 
external galaxy rotation curves \citep[M31 for e.g.][]{bran60}: 
\begin{equation}
\Omega\left(R\right)=\frac{3^{3/2n}v_{max}}{R_{max}\left[1+2\left(R/R_{max}\right)^{n}\right]^{3/2n}}
\label{rceqn}
\end{equation}
The fit is very robust and shows that linear rotation $v(R)$ rises steadily 
from $R=$3-8~kpc ($n=$0.95), peaking at the Solar velocity $v_{max}\sim 
v_0=$255~km~s$^{-1}$ at $R_{max}=$9.3~kpc. Thereafter it declines very gently 
to 235~km~s$^{-1}$ at $R=$20~kpc. Obviously Eqn. \ref{rceqn} cannot capture 
details of non-circular motions in the Inner Galaxy ({}``wiggles'') and the 
sharp discontinuity preceding Perseus Arm objects in the Outer Galaxy, almost 
certainly the signature of a shock (see Sec. \ref{dwtext}). The slope of Eqn. 
\ref{rceqn} is undoubtedly altered by it.

As 21~cm surveys of the MW (e.g. GALFA\ion{H}{i}, ASKAP) evolve to higher 
resolutions and sensitivity so too the measured Galactic rotation curve will 
evolve towards a better sampled, lower scatter version of what we see today. As 
we advance into the next decade, VLBI parallax distances and proper motions to 
more kinematic tracers will sharpen the Outer Galaxy curve and its details 
tremendously \citep{reid09a}. It is imperative that we now begin modelling the 
smooth underlying rotation curve and density wave perturbations separately in a 
multi-parameter model; not an easy task, but we believe the next avenue of 
immediate advancement in MW rotation curve studies. This will also help to 
answer the long-standing question of the asymmetry between the QI and QIV 
rotation curves. Can this be accounted for purely by spiral structure? 

\subsection{Spiral Dynamics: Density Wave Motions in the MW Disc}
\label{dwtext}
The choreographed movement and organization of the old stellar disc into a set 
of nested elliptical orbits creates regions of enhanced stellar density, 
forming a non-axisymmetric gravitational potential $A$ an order of magnitude or 
two smaller than the total potential $V_0$. The pattern in the stars circulates 
as a spiral density wave (hereafter DW) through the disc, and the response of 
the ISM gas is vigorous and complex, with $A(R)/V_0$ as little as 3-5\% 
producing sharp shocks in the gas and forming many secondary and tertiary 
spiral armlets downstream of the potential minimum.
As shown by \cite{robe75} a galaxy's Hubble type and degree of development of 
spiral structure is determined by the strength of shocks in its ISM, which in 
turn is governed by two key DW parameters: the pattern's angular speed of 
rotation $\Omega_p$ and the winding (pitch) angle of the spiral arms $i$. 
$\Omega_p$ is one of the most poorly constrained parameters in all of Galactic 
astronomy. Estimates are based on either stellar kinematic tracers giving 
$\Omega_p=$20-30~km~s$^{-1}$~kpc$^{-1}$ \citep[see][for review]{gerh08}, or on
linking periods of terrestrial climate change with the Sun's passage through 
spiral arms, giving 12-20~km~s$^{-1}$~kpc$^{-1}$ \citep[][]{gies05,sven06}, 
though \citet{over09} calls this link into question. It is not even clear if a 
single constant value for $\Omega_p$ can adequately describe the Milky Way 
spiral. Measurements of $i$ are across the entire range of SBa to SBc galaxies 
\citep[6-20\deg][]{vall08}, probably in part because no distinction is made 
between what is being measured: the DW potential (stellar) arms, shocks 
(\ion{H}{i} self-absorption \& dust), and star-formation arms (\ion{H}{ii} gas and 
OB stars) are all expected to have different $i$ \citep{gitt04}. No conclusive 
confrontation of observations with the non-linear DW (shock) theory has been 
presented so far, and where the MW fits into the spectrum of spiral galaxies 
remains a mystery. Are we more like galaxy M33 (smooth continuous ISM flow; 
poorly defined spiral structure) or M81 (sharp shocks and arms in ISM and many 
secondary gaseous arms in the outer disc)?
\subsubsection{Density Waves in the Inner Galaxy}
We can get an initial idea of the pattern speed in the Inner Galaxy from the 
recent GLIMPSE results \citep[Figure 14 in][]{chur09}. Star counts 
towards the Scutum and Centaurus arms show peaks, whereas none are seen towards 
the Sagittarius \& Norma arm tangencies, suggesting that these two arms are 
predominantly ISM concentrations. Theory predicts such secondary arms form from 
the gas' dynamical response to a principal 2-arm stellar DW. A family of 12 
hydrodynamics galaxies are built with combinations of $\Omega_p=$12,16,20 and 
$i=$8,10,12,14\deg, spanning today's range of contentious values for these 
parameters. Surface density images of these simulations are shown in online 
figures at www.novell.brandonu.ca/$\sim$fostert/DWModels.pdf. In them, the 
Sun is the bright dot located at $R_0=$7.8~kpc; its phase with respect to the 
pattern is set by the parallax distance to W3OH ($\ell=$134\deg, $r=$2~kpc), 
the location of the Perseus arm shock \citep[][]{xu06}. All models form a 4-arm 
ISM pattern beyond some radius which is relatively insensitive to pitch angle 
($R\gtrsim$11~kpc, 8.5~kpc \& 6~kpc for $\Omega_p=$12, 16 \& 
20~km~s$^{-1}$~kpc$^{-1}$ respectively). The observation of gaseous arms deep 
in the Inner Galaxy at $R\sim$3-5~kpc (beyond the bar) suggests then that 
$\Omega_p\gtrsim$20~km~s$^{-1}$~kpc$^{-1}$ there.

\begin{figure}[ht]
\centerline{
\includegraphics[scale=0.21,clip=]{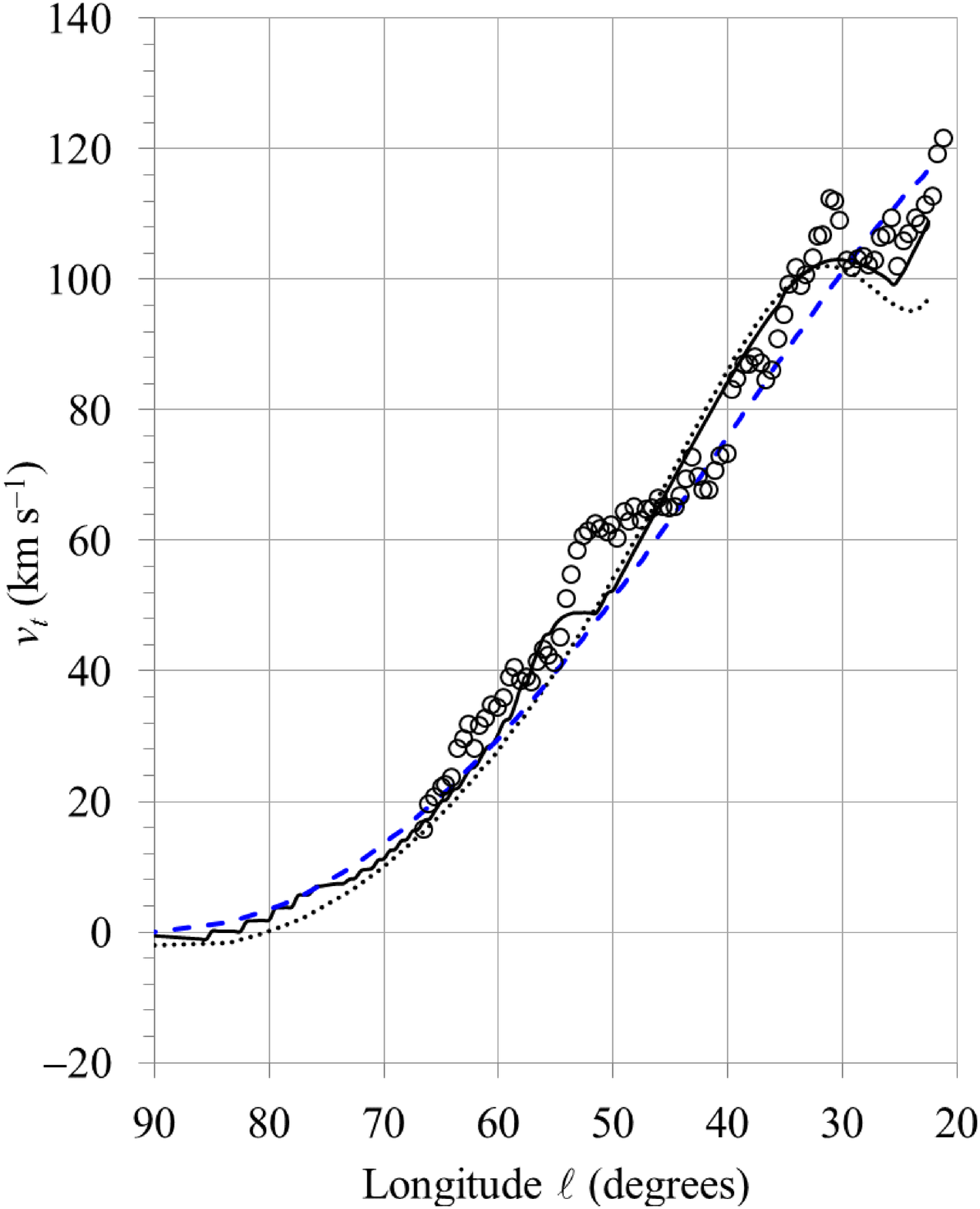}
\includegraphics[scale=0.25,clip=]{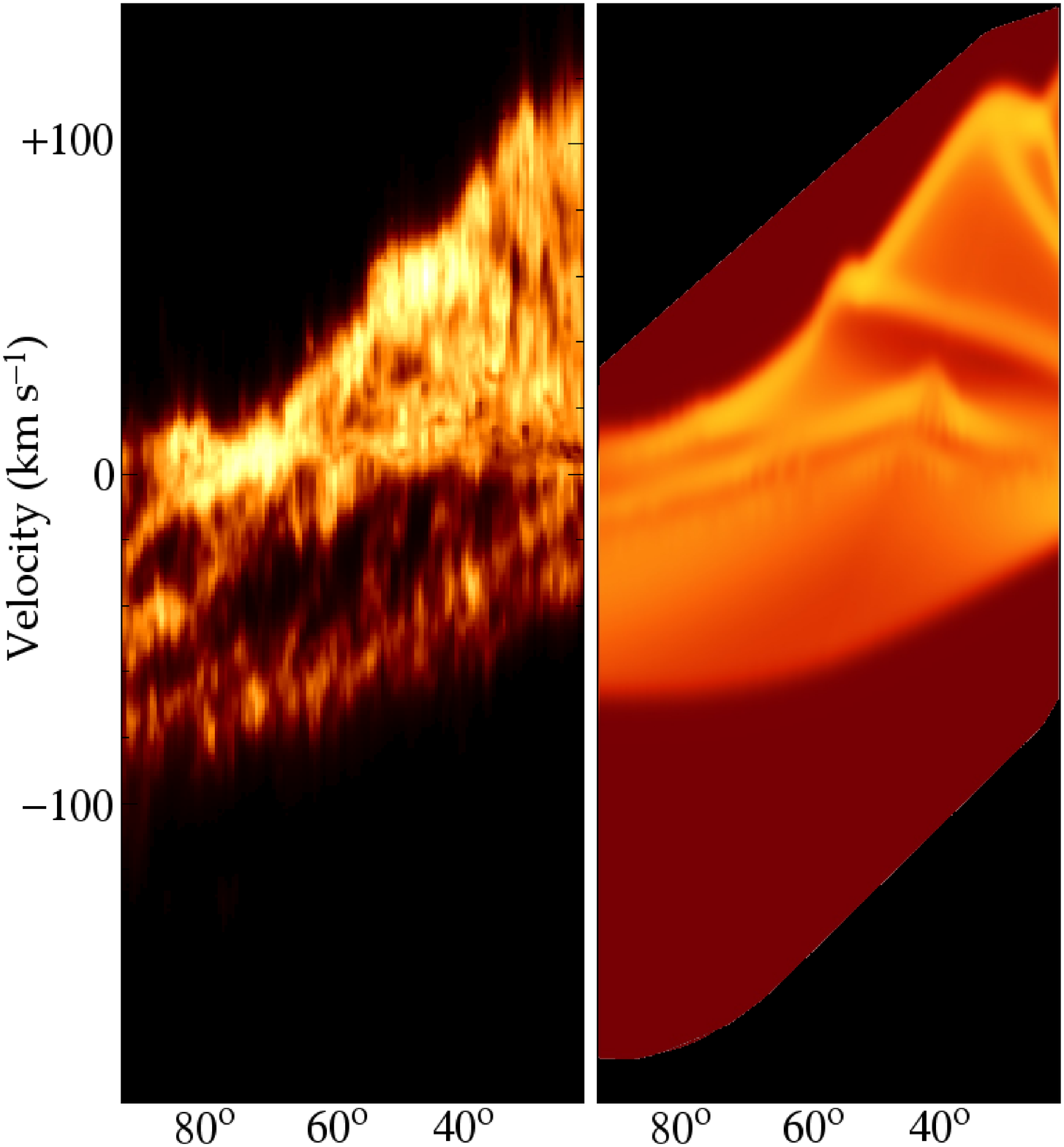}}
\caption{(Left) Inner Galaxy (QI) terminal velocities $v_t$ predicted from 
circular rotation (dashed), linear DW theory (dotted) and non-linear theory 
(solid), compared to measured ones from \citet{levi08} (circles). (Right) 
Synthetic $T_b\left(\ell,~v_{LOS}\right)$ diagram (right) compared to the 
Leiden \ion{H}{i} survey (left) for $\ell=$22-90\deg. Both panels are 
calculated from the same galaxy model $\Omega_p=$20~km~s$^{-1}$~kpc$^{-1}$ and 
$i=$14\deg.
}
\label{ingaldw}
\end{figure}

The best evidence for density waves in the Inner Galaxy are the wavy 
irregularities in \ion{H}{i} terminal velocities ($\sim\pm$10~km~s$^{-1}$) with 
Galactic longitude (see Fig. \ref{ingaldw}). Here as well spiral structure, 
inextricably linked with spiral dynamics, is staring us in the face and crying 
out to be properly (quantitatively) modelled with theory in a parameter fitting 
analysis. However, Fig. \ref{ingaldw} shows that $v_t$ is too poorly resolved, 
and that the signatures of linear and shock theories are too similar to 
conclusively choose between them with the current data. It is not clear whether 
more careful measurements of $v_t$ will unambiguously resolve the sharp cusps 
in $v_t$ that shock theory predicts, one of the only features at the tangent 
point distinguishing it from the linear version. Nonetheless, the use of 
high-resolution 21~cm data as illustrated by \citet{mccl07} is the correct way 
to begin to resolve DWs in the Inner Galaxy.

The most natural way to model DWs is by comparing theoretical 
$T_B\left(\ell,v_{LOS}\right)$ diagrams to observed ones, as theory predicts
both ingredients that shape these diagrams: the (perturbed) density and total 
velocity (assuming constant spin temperature). Qualitative comparisons of the 
$\ell,v_{LOS}$ diagram were made very early by \cite{burt71,robe72,sawa78} and 
others, and once again, it was found that linear and shock version of DW theory 
produce very similar diagrams with only subtle differences \citep[][]{sawa78}. 
Fig. \ref{ingaldw} (right) shows the observed QI $\ell,v_{LOS}$  diagram 
(Leiden \ion{H}{i} data) compared to that simulated from the model galaxy with 
$\Omega_p=$20~km~s$^{-1}$~kpc$^{-1}$, $i=$14\deg, the model we find that best 
fits the Leiden observations in a simple 
$\chi^2=\Sigma\left(T_B^{obs}-T_B^{mod}\right)^2$ minimization.

\subsubsection{Density Waves in the Outer Galaxy}
\begin{figure}[ht]
\centerline{
\includegraphics[scale=0.15,clip=]{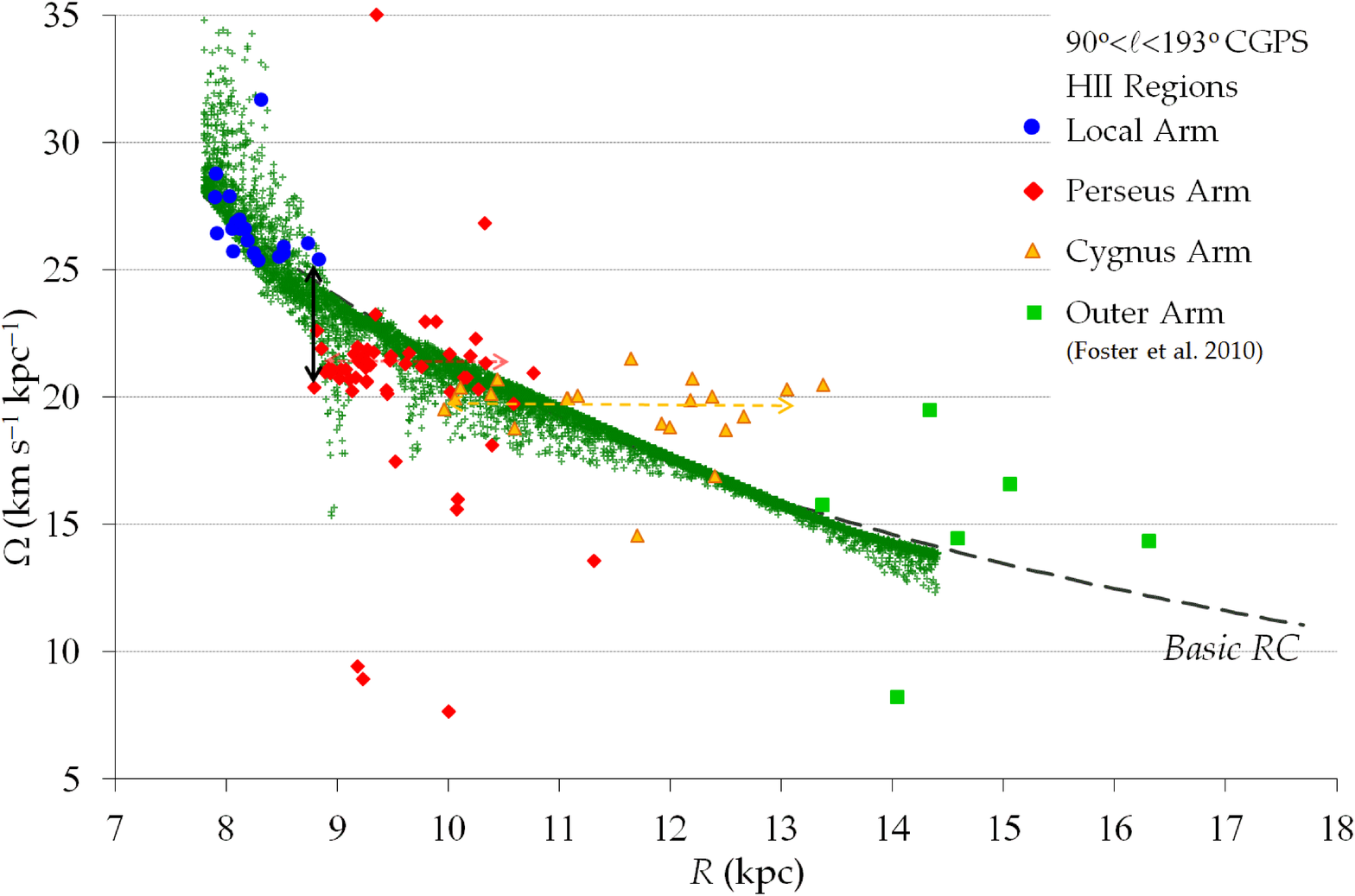}
\includegraphics[scale=0.15,clip=]{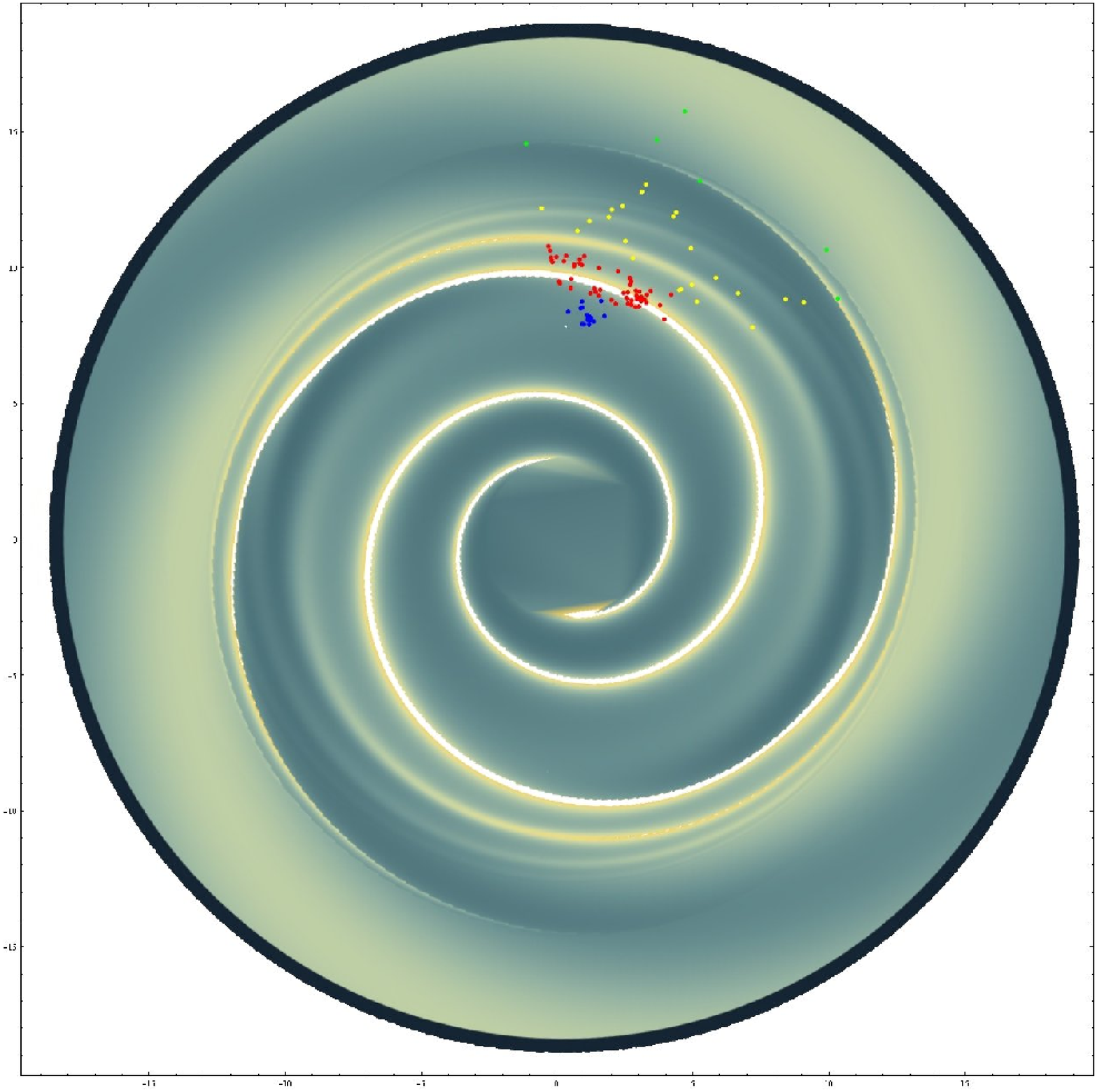}}
\caption{(Left) Outer Galaxy $\Omega(R)$ observed with 100 \ion{H}{ii} regions 
in the CGPS, plotted with the simulated rotation curve observed in a synthetic 
galaxy (model $\Omega_p=$16~km~s$^{-1}$~kpc$^{-1}$, $i=$12\deg). (Right) Surface density map of model 
$\Omega_p=$12~km~s$^{-1}$~kpc$^{-1}$, $i=$12\deg with 100 CGPS \ion{H}{ii} 
regions overlaid \textit{not fitted}. See 
http://www.novell.brandonu.ca/$\sim$fostert/DWModels.pdf for larger figures.}
\label{outgaldw}
\end{figure}

In the Outer Galaxy (QII \& QIII) the velocity anomaly associated with Perseus 
spiral arm \ion{H}{ii} regions has been long known, and is seen in published 
rotation curves today between $R=$9-11~kpc as a cluster of objects underneath 
the smooth $v(R)$ (see Fig. \ref{igpsrc}. The explanation in terms 
of the shock was proposed early on by \citet{robe72}, but the idea was never 
really explored further from there, probably because the scatter in rotation 
curves based on \ion{H}{ii} regions prevents a positive identification of a 
sharp feature like a shock. However, the Outer Galaxy rotation curve traced 
with CGPS \ion{H}{ii} regions clearly shows this velocity discontinuity between 
the local \& Perseus arm objects at $R=$8.9~kpc (see Fig. \ref{outgaldw}, 
left). To compare this to, $\Omega(R)$ is simulated from our family of 
synthetic galaxies by picking any point a known distance from the Sun's 
location and assuming its entire LOS velocity projection is described by Eqn. 
\ref{fundeqn}. The results of picking thousands of these points show the type 
of rotation curve one might measure for such a galaxy. Fig. \ref{outgaldw} 
(left panel) is the simulated rotation curve from model 
$\Omega_p=$16~km~s$^{-1}$~kpc$^{-1}$, $i=$12\deg indeed showing that a shock 
would appear as a discontinuity in $\Omega$ extending under the basic 
rotation curve at $R=$8.9~kpc, just as is seen in the Perseus arm \ion{H}{ii} 
regions. We have found that only simulations with $\Omega_p=$16 place the shock 
exactly at their location; $\Omega_p=$12 models place the shock well ahead of 
the \ion{H}{ii} regions, and $\Omega_p=$20 models do not form shocks at all 
beyond the Solar radius.

We distinguish a group of more distant \ion{H}{ii} regions at a constant 
$\Omega=$20\,km\,s$^{-1}$\,kpc$^{-1}$ which trace a second optical arm some 
$\sim$2~kpc beyond Perseus. The presence of second optical arm beyond the 
Perseus arm was proposed by \citet{kime89}, and stars in it traced by 
\citet{negu03} who refer to it as the {}``Cygnus arm''. In our sample some 20 
objects are distinguished from the Perseus arm members by not only 
their distinctly larger distances ($R_{ave}\sim$11~kpc), but by 1) their 
relatively high latitudes (mean 2\fdg5) compared to the Perseus arm, 2) their 
correspondence with an \ion{H}{i} concentration similarly higher up in the 
distant warped midplane (see Fig. \ref{rollhi}), and 3) their appearance in a 
larger {}``kinematic ring'' 
$(\Omega-\Omega_0)\propto\left|v_{LOS}\textrm{/sin}\left(\ell\right)\right|$. 
In our simulated rotation curve Cygnus arm objects appear behind a secondary 
shock beyond the Perseus arm (Fig. \ref{outgaldw}). The presence of both 
star-forming arms at $R>R_0$ has implications for the pattern speed $\Omega_p$, 
if we are to assume that the \ion{H}{ii} regions that trace them form from 
shocks. The right panel shows a birds-eye view of model 
$\Omega_p=$12~km~s$^{-1}$~kpc$^{-1}$, $i=$12\deg. This model generates multiple 
strong secondary shocks behind the primary one in the Perseus arm (at 
$R=$8.9~kpc) that explains the presence of \ion{H}{ii} regions in the Cygnus 
and Outer arms (yellow and green respectively). The $\Omega_p=$16 models do not 
generate shocks beyond the Perseus arm shock, and $\Omega_p=$20 models show no 
strong shocks form at all beyond $R=R_0$, making the presence of any massive 
star-formation regions outside the Solar circle not as easily explained.

\begin{figure}[ht]
\centerline{
\includegraphics[scale=0.34,clip=]{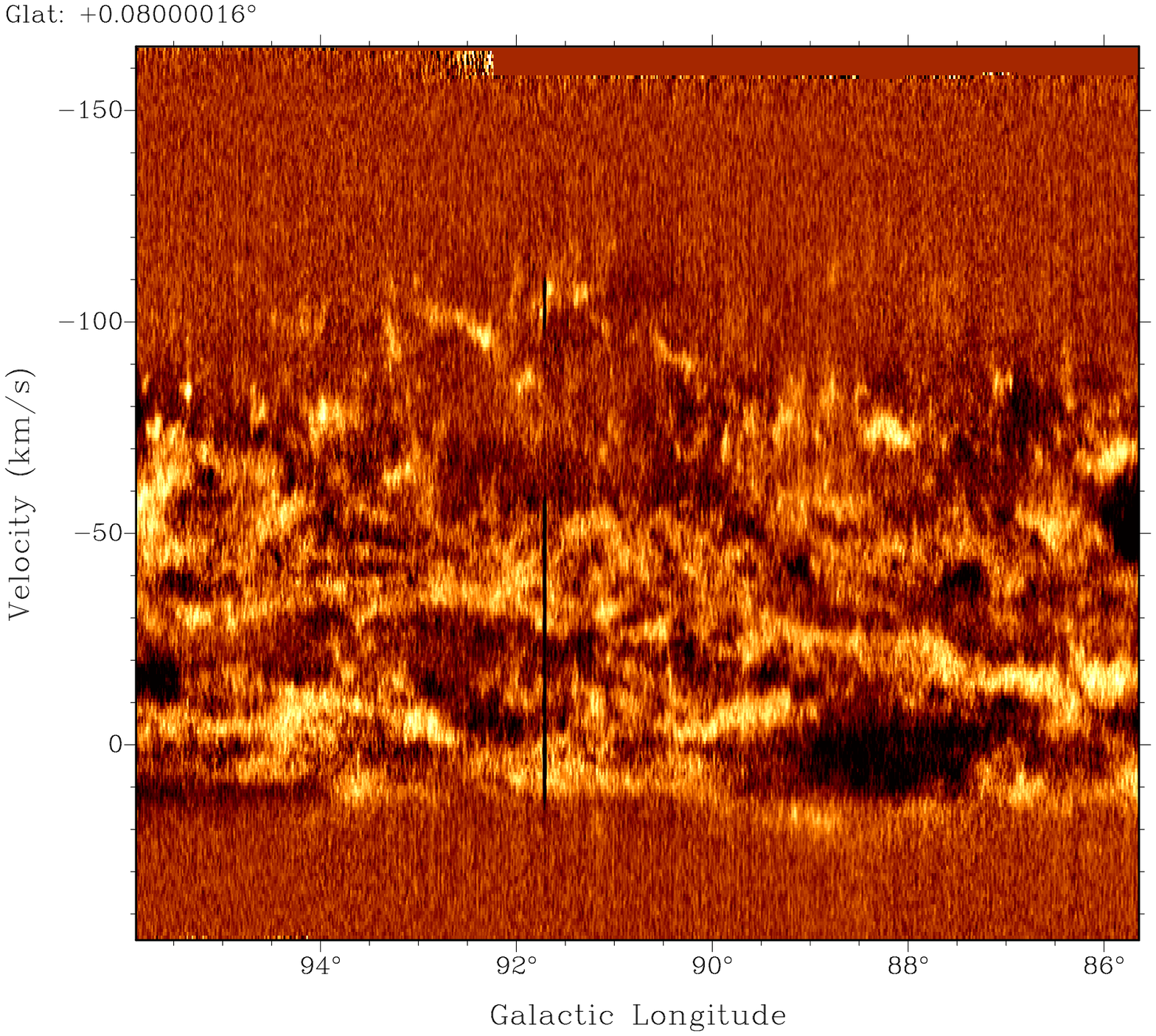}
\includegraphics[scale=0.34,clip=]{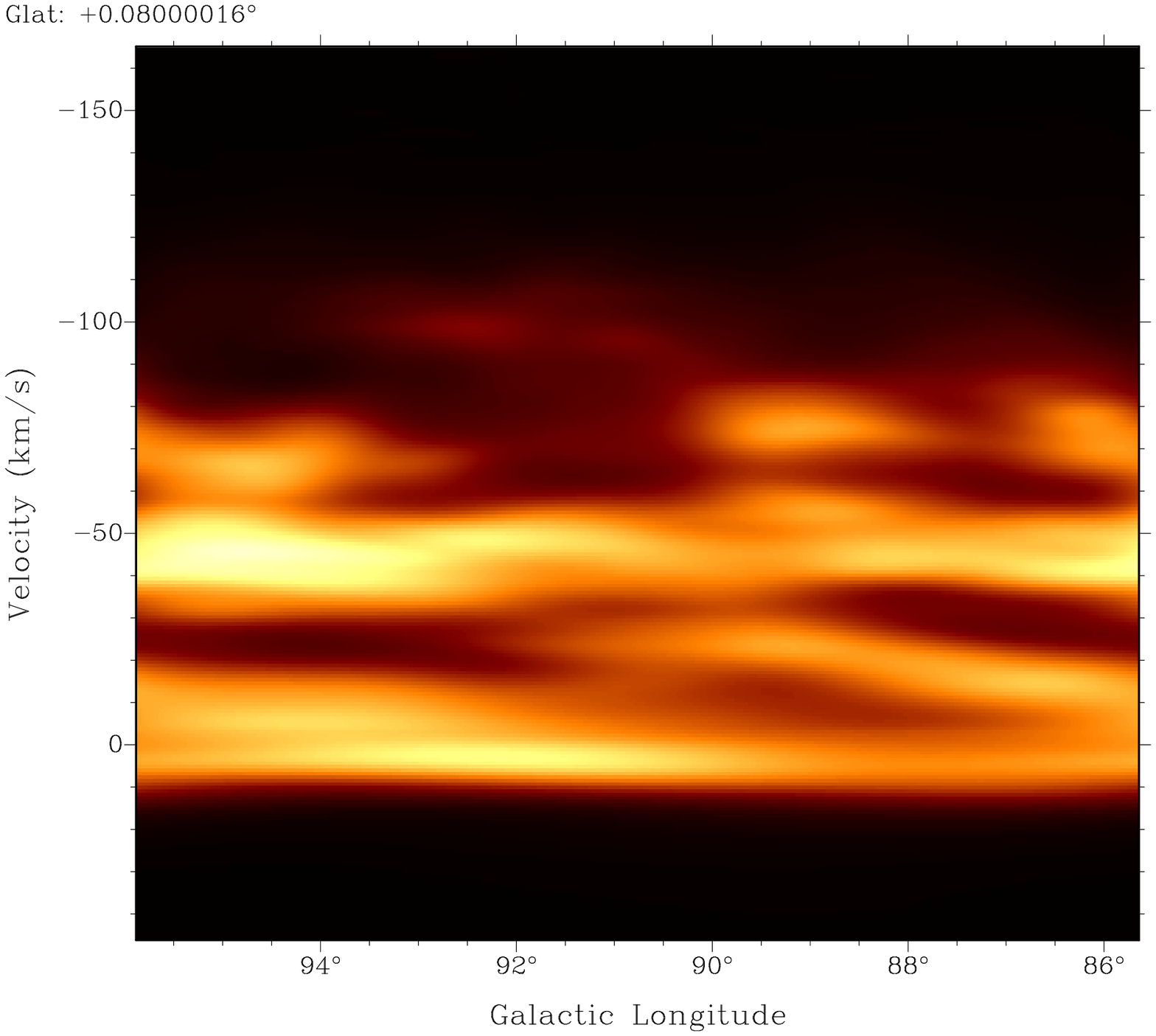}}
\caption{A 10\deg~wide $T_B\left(\ell,v_{LOS}\right)$ plot from CGPS \ion{H}{i} 
data separated into {}``cloud'' (left) and {}``intercloud medium'' (right) maps 
using a spatial filter. Very little coherent large-scale 
structure remains in the cloud component, in contrast to the intercloud map at
right which is readily compared to DW models.}
\label{filteredcgps}
\end{figure}

It is likely then that the MW has a radially falling spiral pattern speed of 
the kind seen in some external spirals \citep{meid09}, with $\Omega_p\gtrsim$20 
from the bar's edge to $R\sim R_0$, dropping to $\Omega_p\sim$16 at 
$R\sim$9~kpc and to $\Omega_p\sim$12~km~s$^{-1}$~kpc$^{-1}$ at 
$R\gtrsim$11~kpc. Admittedly this conclusion is drawn from somewhat simple 
comparisons made for the sake of this review, but we believe demonstrate what 
is possible to discover in confronting DW theory with kinematics of the Outer 
Galaxy. Large-scale comparisons of this kind with new observations have today 
fallen behind the ever-increasing number of theoretical simulations. 
The power in new high-resolution 21~cm surveys will be their ability to resolve 
the smooth intercloud medium between the clouds, allowing \ion{H}{i} cubes that 
show a less-confused view of large-scale Galactic motions like rotation, 
density wave streaming and turbulence to be created. \citet{rohl74} recognized 
this and applied a minimum-brightness boxcar type filter to isolate the 
large-scale \ion{H}{i} {}``stratum'' from the {}``clouds''. \citet{levi06} use 
an unsharp mask on LAB \ion{H}{i} data to enhance the visibility of spiral 
structure in surface density maps. A similar more advanced filter based on the 
background separation technique in \citet{koth02} is demonstrated with CGPS 
data in Fig. \ref{filteredcgps}. 
With any filter it is important to use high angular resolution observations in 
order to resolve the narrow gaps between clouds. Creating new expressions of 
the neutral ISM images will ultimately resolve the subtle differences between 
density wave theories.

\subsection{Rolling Motions}
\label{rolltext}
Spiral arms in most directions in the MW exhibit strange velocity gradients 
perpendicular to the plane. These were first observed in the 21~cm line by 
Oort \citeyearpar{oort62} and coined {}``rolling motions''. The term comes from 
the arms' tilted appearance in \ion{H}{i} latitude-velocity slices 
$T_B\left(b,v_{LOS}\right)$; at positive latitudes the $T_B$ centroid of the 
arm appears at higher negative velocities, whereas down through the equator and 
below ($b<0$) the centroid smoothly shifts to more positive velocities. Rolling 
is also observed in the molecular ISM \citep[$^{12}$CO,][]{wout81}, and in 
local young stars \citep{bros84}. 
Discrete objects like \ion{H}{ii} regions also {}``roll'', exhibiting the same 
overall gradient across the plane as the \ion{H}{i} arms do \citep[][]{fost10}, 
and are therefore likely tracing the same motions. Typical gradients are 
$dv\textrm{/}db\sim$2-4~km~s$^{-1}$ per degree of latitude, or about 
$dv\textrm{/}dz\sim$20~km~s$^{-1}$ per kpc in the Perseus arm. Rolling motions 
have not yet been observed in external galaxies.

\begin{figure}[ht]
\centerline{
\includegraphics[scale=0.31,clip=]{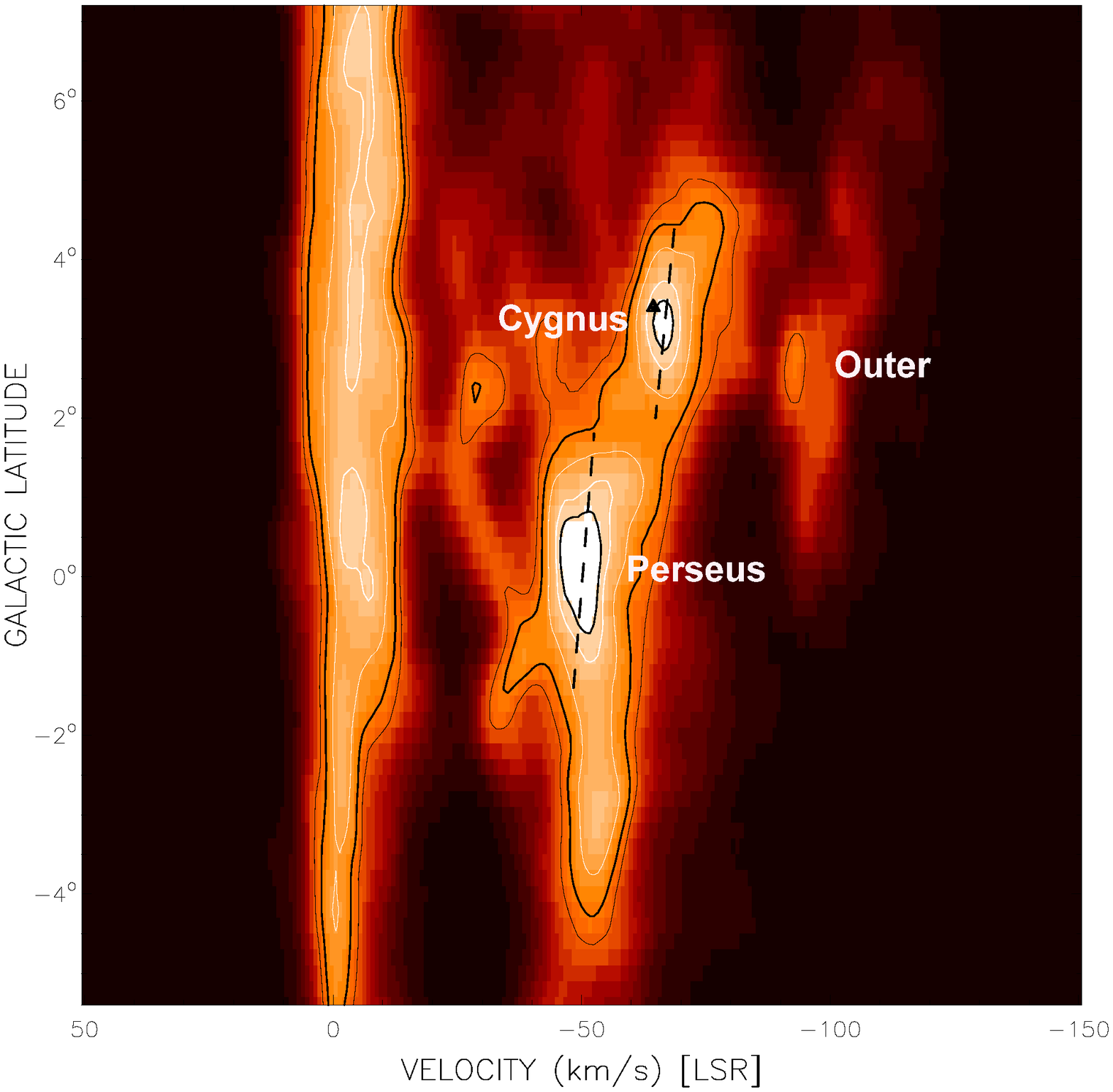}
\includegraphics[scale=0.29,clip=]{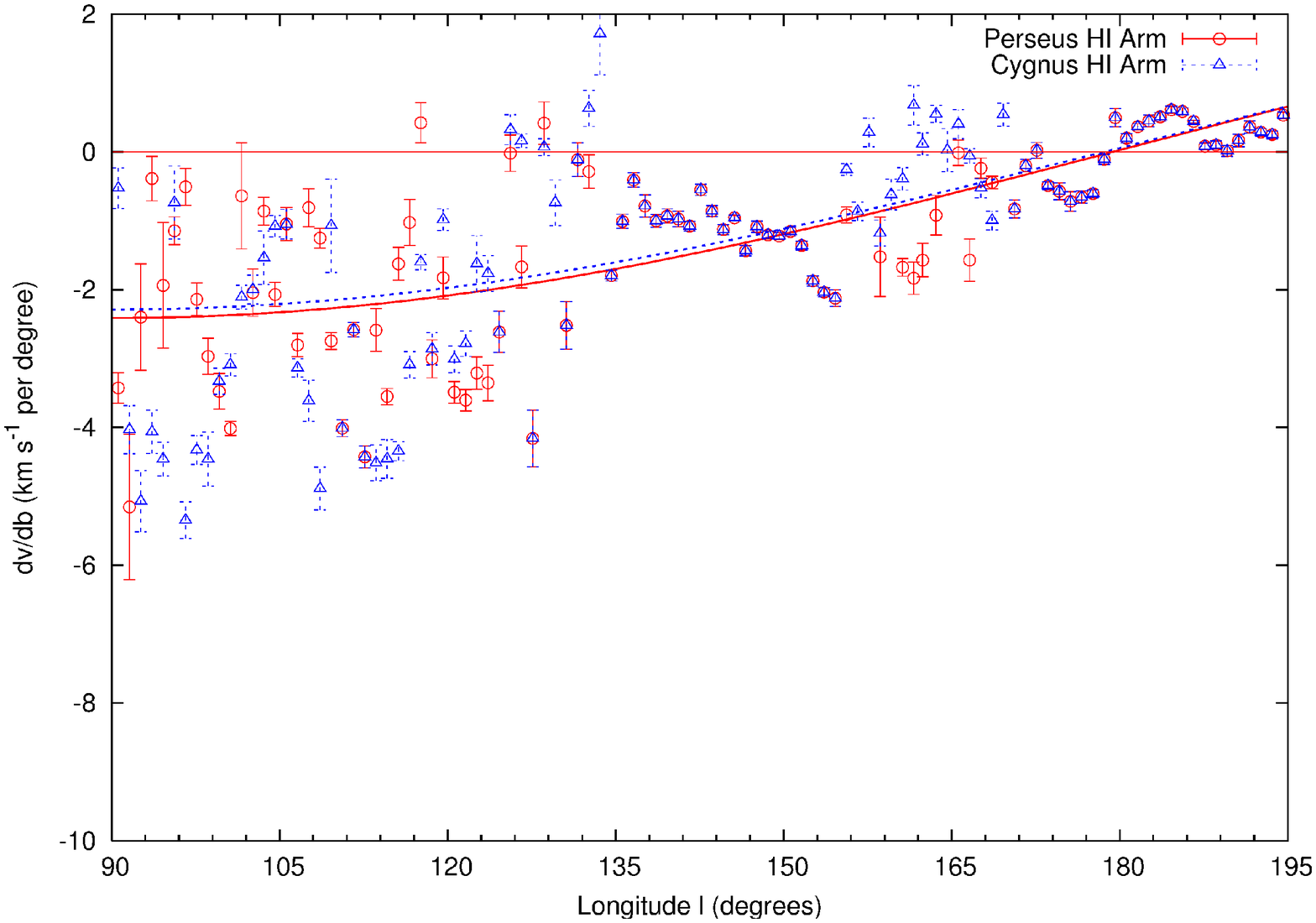}}
\caption{(Left) $T_B\left(b,~v_{LOS}\right)$ slice from the DRAO 26-metre 
\ion{H}{i} survey \citep{higg00}, showing the Perseus ($v_{LOS}\sim 
-$50~km~s$^{-1},~b\sim$0\deg), the Cygnus ($-$70~km~s$^{-1},+$3\deg) and the 
Outer \ion{H}{i} arm ($-$90~km~s$^{-1},+$2.8\deg). The triangle marks the 
$b,~v_{LOS}$ of Sh2-141, seen within the \ion{H}{i} concentration that 
\citet{fost10} identify with the Cygnus arm. (Right) Gradients measured across 
QII for both arms. Functions 
$dv(\ell)\textrm{/}db=-$2.44sin$(\ell)+$0.03 and $-$2.35sin$(\ell)+$0.06 fitted 
to the Perseus (solid line) and Cygnus arms (dotted line) respectively are 
drawn.}
\label{rollhi}
\end{figure}

\citet{yuan73} first described rolling as apparent {}``pseudo-motions'' created 
by our top-down view of the arms caused by their tilt towards up (due to the 
warp of the midplane), and differential circular rotation. For example, in QII 
the upward warp causes a given $T_B$ contour of an arm at higher latitudes to 
be observed at a greater distance (and thus at a more negative velocity) than 
the same contour at lower latitudes. This geometric effect is subtle, and is 
reproducable in 3D models of the Galactic disk that include the warp 
\citep[e.g.][]{fost06}, however, a modern flat rotation curve and figure for 
the warp produces smaller apparent gradients ($<$1~km~s$^{-1}$ per degree) than 
observed. This was also shown by \citet{feit85}, opening the possibility that 
rolling motions are partly physically real. They \citeyearpar{feit86} propose 
tumbling pockets of \ion{H}{i} gas falling from high in Galactic Fountains, 
themselves caused by active star formation in the arm's middle. 

The \citeauthor{yuan73} model seems to be supported by observations in the 
Outer Galaxy. If rolling is entirely due to observed differences in circular 
motion (including rotation) with latitude, then gradients in the arms should 
disappear in the cardinal direction $\ell=$180\deg. This is observed for both 
the Perseus \& Cygnus \ion{H}{i} arms (see Fig. \ref{rollhi}). However, the 
latitudes $b$ of Perseus 
arm \ion{H}{ii} regions are constant ($b_{ave}=$0\fdg1) with 
increasing distance $R$, yet these objects also {}``roll'' with the \ion{H}{i} 
arm. Rolling in Perseus must therefore be entirely physical and not a viewing 
effect. The magnitude of the rolling gradient in Cygnus arm objects is the 
same as in Perseus, so rolling in the Cygnus arm could also have essentially 
nothing to do with its displacement above the midplane. However \ion{H}{ii} 
regions associated with this arm \textit{do} show a relationship of $R$ with 
latitude $b$, just as one would expect if the warp were involved!

As the reader can see there are more inconclusions than conclusions. How much 
of the observed gradients in the MW are real, and how much apparent? A good 3D 
MW model that includes the warp fitted to the observations will be necessary to 
separate real from apparent rolling motions, after which the fountain theory 
can be further tested or other explanations sought. Are rolling motions unique 
to the MW, and therefore likely illusions caused by our viewing point? Future 
21~cm observations of external galaxies will tell us. The EVLA is the only 
instrument with resolution $\leq$1~km~s$^{-1}$ and $\sim$1\arcsec on the sky 
sufficient to scrutinize external spiral arms with 50~pc spatial resolution at 
D=10~Mpc. Nearby galaxies in the THINGS survey (see this volume) would be good 
starters, and in the future ASKAP will be able to extend our observations. 

\section{Galactic Structure}
\label{galstructtext}
An executive summary of the MW might go something like this: the MW is a barred 
spiral galaxy with a short boxy bulge in its centre, and a stellar bar about 
twice as long. The old stars form a principally 2-arm DW pattern in the Inner 
Galaxy, giving rise to at least 4 ISM arms that are active in star formation 
from $R\sim$4-14~kpc. The 4-arm pattern winds through most of the Inner and 
all through the Outer Galaxy with the same pitch angles of 
$i\sim$12\deg, and the arms are connected by minor branches and spurs 
throughout, including one that the Sun currently resides in. Inside the Solar 
circle the \ion{H}{i} disc is remarkably flat, while outside the disc is 
twisted nearly along the Sun-centre line into a gentle warp above the 
midplane in one hemisphere and below it in the other.
\subsection{Milky Way Spiral Structure: Methods and Results}
\label{spiralstructtext}
\begin{figure}[ht]
\centerline{
\includegraphics[scale=0.25]{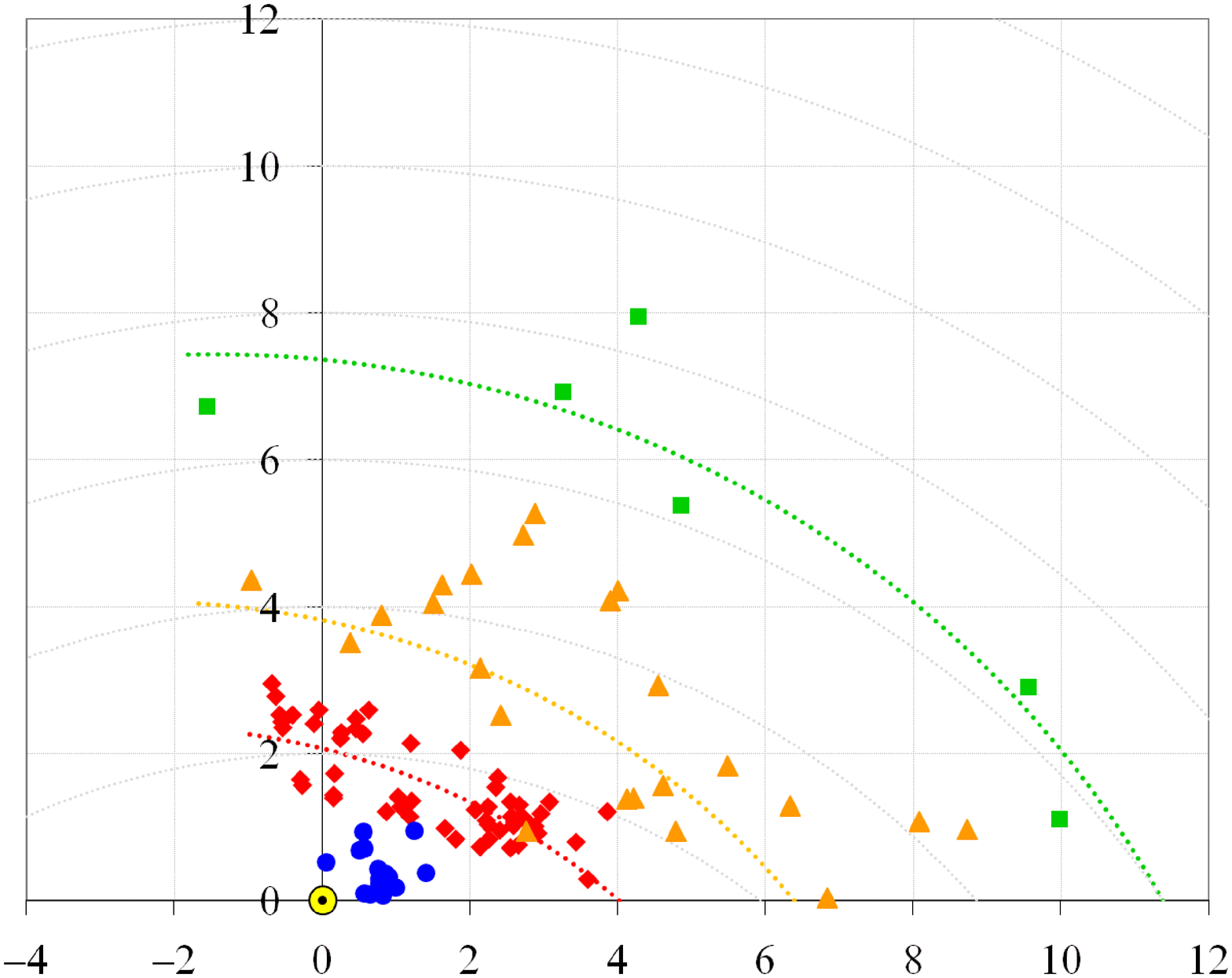}
\includegraphics[scale=0.42]{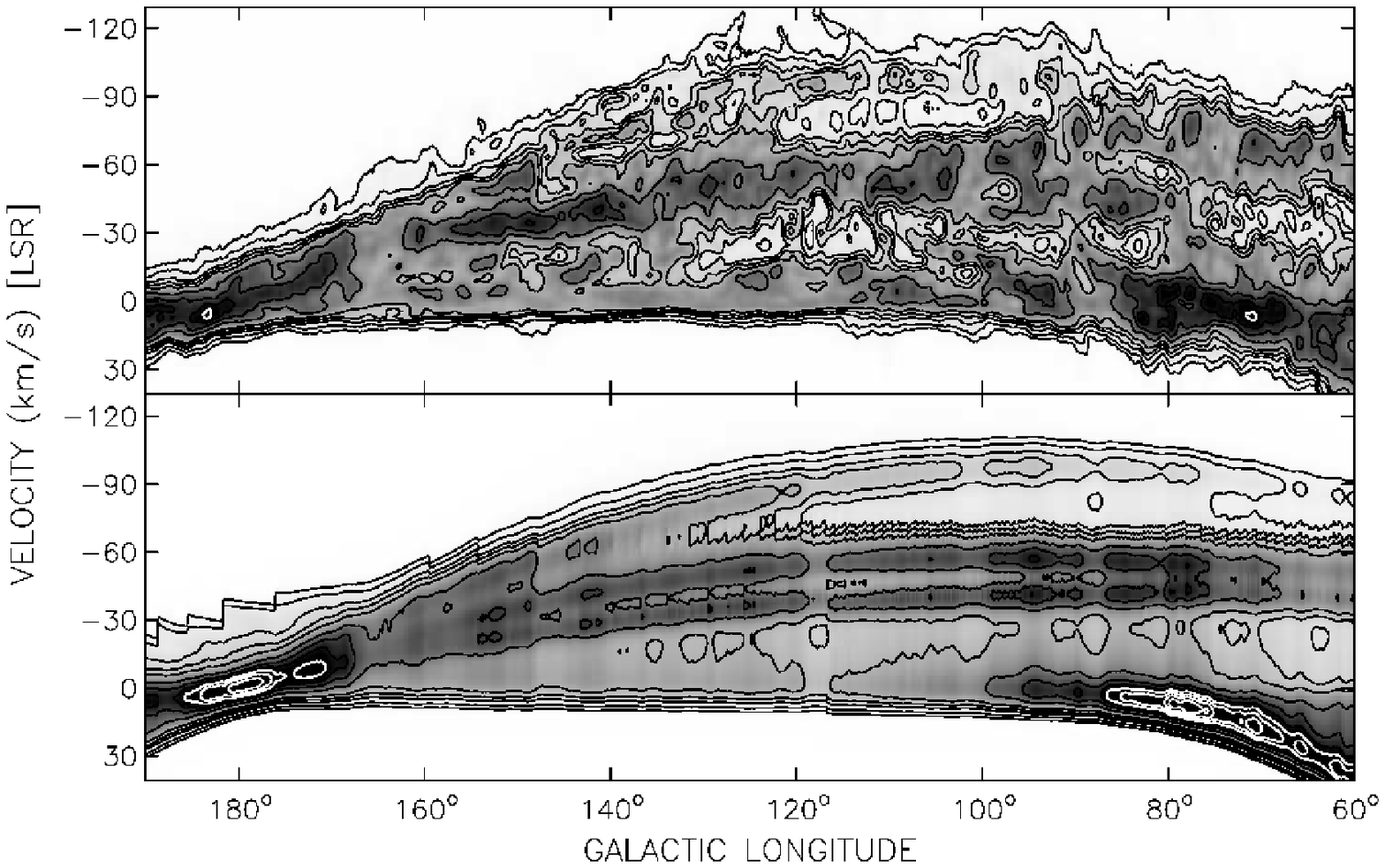}}
\caption{(Left) Spiral structure in QII and part of QIII from CGPS \ion{H}{ii} 
regions. Fitted pitch angles are: Perseus arm (red diamonds) 
$i=$12.0$\pm$2.6\deg; Cygnus arm (yellow triangles) 12.6$\pm$4.1\deg, and Outer 
arm (green squares) 5.6$\pm$3.9\deg. (Right) $T_B\left(\ell,v_{LOS}\right)$ 
plot at $b=$0\deg from the 26~m \ion{H}{i} survey by \citet{higg00} 
compared to the model of \citet{fost06} fitted to it quantitatively.}
\label{pitch}
\end{figure}

Mapping the spiral structure of the Milky Way historically is still an exciting 
challenge that continues to this day. The historic 21~cm map by Kerr-Westerhout 
is based on 1) \textit{velocimetric deconvolution of the observed 
$T_B\left(\ell,v_{LOS}\right)$ brightness distribution} using circular rotation 
and a constant spin temperature. \citet{levi06} recreate only the Outer Galaxy 
this way using the LAB \ion{H}{i} data, and a minor refinement to the velocity 
field. Although a bit better than the historic map, the same distorted picture 
emerges with pitch angles $i=$21-25\deg, differing wildly from those more 
directly measured with kinematic tracers. These maps demonstrate that 
\textit{\ion{H}{i} is a poor tracer of Galactic spiral structure when used like 
this}. Deconvolutions based on the \citet{dame01} CO survey are clearer 
\citep[e.g.][find a more reasonable $i=$12-15\deg]{naka06}. \citet{pohl08} use 
a non-circular velocity field based on a hydrodynamics (HD) galaxy simulation, 
itself fitted to the observed NIR luminosity distribution. The resulting Inner 
Galaxy map is more detailed than any previous one, showing a distinct bar in 
the centre angled to the Sun-centre line by $\phi=$20\deg, and two arms 
emanating from the bar's ends at $R\sim$3.5~kpc and winding outward by 
$i\sim$11\fdg5. 
Thus far no study has applied such a model to deconvolution of the observed 
\ion{H}{i} distribution, though the approach is promising as shown by 
\citet{gome06} in reconstruction of observed simulations. The more flexible 
approach of 2) \textit{modelling the observed $T_B\left(\ell,b,v\right)$ 
distribution} is demonstrated in Fig. \ref{pitch} (right). An empirical DW 
model \citep{fost06} created the synthetic $\ell-v$ plot at bottom, and 12 
adjustable parameters were varied in a proper least-squares fit to the data 
(top). Only two DW arms are modelled. Tight winding angles of $i=$6\deg are 
found, primarily because the model treats the Cygnus and Outer arms as one. 
The computational cost of this multi-parameter model-fitting approach is not 
prohibitive as it is for more realistic MHD models \citep{gome06}. Empirical 
models can also include components that have been difficult to incorporate in 
smoothed particle hydrodynamics models, such as the warp, magnetic fields, and 
rolling motions.

The paradigmatic 4-arm map of \citet{geor76} is based on 3) \textit{direct 
spatial mapping} using 268 \ion{H}{ii} regions with photometric distances 
and/or kinematic distances based on RRL and H$\alpha$ velocities. The 
\citeauthor{geor76} map has achieved some fame (cited at least 550 times) and 
some infamy. It is the starting point for the also-famous electron density map 
by \citet{tc93}, and the dust emission model by \citet{drim01}. Compared to 
modern maps the pattern in \citeauthor{geor76} is unusually (and worringly) 
clean, and has prompted many to propose the MW is among the 10\% of spirals 
with a so-called {}``grand design''. While it is generally accepted that there 
is indeed a coherent spiral pattern to be mapped,
\ion{H}{ii} regions and giant molecular clouds in the MW sport a truly messy 
pattern \citep[c.f.][]{pala04,hou09}, revealed by {}``photoshopping'' out the 
fitted arms drawn on published \ion{H}{ii} region maps. 
One wonders if \ion{H}{ii} regions will ever reveal a 
{}``grand design'' spiral structure in the MW, if one exists at all. Recent 
\ion{H}{ii} region maps by \citet{russ07,hou09} fit 2-4 logarithmic or variable 
pitch angle spiral segments, with 4 arms at $i=$10-13\deg always fitting better 
than two with tighter $i=$6-8\deg windings. Of course, any model with twice the 
free parameters \textit{should} fit better, but 4-arm models also reproduce 
known arm tangency directions. Within the likely uncertainties of the fits 
($\pm$20-30\%; see below), 4-arm models show the same pitch angle arm to arm 
and negligible variation of $i$ with radius (this however is not clear 
as few papers offer proper uncertainties on any fitted spiral arm parameters!). 
A least-squares fit to CGPS \ion{H}{ii} regions weighted by uncertainties in 
both coordinates ln$R$ \& $\phi$ yields $i=$12.0$\pm$2.6\deg and 
12.6$\pm$4.1\deg for the two major star-forming arms in QII (see Fig. 
\ref{pitch}).

Another method, 4) \textit{fitting models to Galactic features observed on the 
sky} (e.g. arm tangency directions) yields information on the number of arms as 
well as large symmetric structures like the disc scale length, thickness and 
flaring, and the warp. An example is the model of \citet{drim01} based on COBE 
observations of NIR (stars) and FIR (warm dust) emission. \citeauthor{drim01} 
intriguingly find that an $m=$2 arm pattern dominates the nonaxisymmetric NIR 
emission. This is supported by Inner Galaxy star counts from the GLIMPSE 
survey, which show no enhancement in star numbers towards the expected tangency 
of the Sagittarius arm $\sim$45\deg$\leq\ell\leq$51\deg \citep[Fig. 14 
in][]{chur09}. A visualization of the MW constrained with GLIMPSE, \ion{H}{i} 
\& CO Galactic plane surveys, and other observations \citep[Fig. 15][]{chur09} 
shows the Perseus and Scutum-Centaurus arms are the principal stellar arms, 
while Sagittarius and Norma are two star-poor ISM arms. The Norma arm wraps 
around from its origin at $R=$4~kpc to become the {}``Outer arm'' in QII \& 
QIII, and then is identified with the distant \ion{H}{i} arm of \citet{mccl04} 
in QIII \& QIV. 

Comparison of the GLIMPSE map with the recent synthesis of MW spiral structure 
by \citet{vall08}, and with CGPS \ion{H}{ii} regions, shows some fundamental 
differences. \citet{vall08} draw the Sagittarius and Norma arms 
attached to the bar's ends, whereas the Scutum-Crux and Perseus arms first 
appear outside the zone of the bar. In the Outer Galaxy the winding angle and 
distance to the Perseus arm agree extremely well in all three maps. 
\citet{vall08} show the Cygnus arm as the Outer Galaxy extension of the Norma arm, however, it appears 
much farther from the Sun (6-7~kpc between $\ell=$130-150\deg) than \ion{H}{ii} 
regions and stars \citep[e.g.][and Fig. \ref{pitch}, this work]{negu03} show it 
to be (4-5~kpc). It is possible that its distance in \citeauthor{vall08} is 
kinematically estimated and overly large due to the presence of a second shock 
beyond the Perseus arm in QII. The Cygnus arm is missing altogether from the 
GLIMPSE map, but would appear halfway between the Perseus and Outer arms on 
there. Given the pervasive confusion in the literature between it and the 
Perseus arm, this is not surprising.

MW spiral structure is then far from a closed subject, and as pointed out by 
\citeauthor{chur09} is deserving of much more attention from the 
community. In particular the relationship between the Inner and Outer Galaxy 
patterns needs to be clarified. How do the stellar and ISM arms from GLIMPSE 
relate to the Outer Galaxy pattern? As computational speeds increase, new 
attempts at determining MW spiral structure in \ion{H}{i} by modelling and 
comparisons with DW theory, and deconvolution with self-consistent dynamical 
models may be profitable. As more VLBI parallax distances can be measured 
towards fainter masers \citep[][]{reid09a}, the precise pattern (or lack 
thereof!) traced out by star-forming regions will become clearer in the decade 
to come. 


\subsection{The Warp}
The pioneering 21~cm observations of the MW disc in the 1950's showed the 
remarkable flatness of the Inner Galaxy disc, and in 
dramatic contrast, the warp of the Outer \ion{H}{i} layer. A warped midplane is 
a commonality shared by most galaxies whose \ion{H}{i} extends past their 
optical discs. Our neighbors M31 and M33 both exhibit warps 
\citep[e.g.][]{chem09}. The warp remains the least understood large-scale 
component of MW structure. It begins at $R_{max}\sim1.5R_0$ inside the stellar 
disc. The Sun-centre line also happens to nearly bisect the warped halves of 
the \ion{H}{i} disc, with the half in the Northern sky much more pronounced in 
amplitude. In QI \& QII, \ion{H}{i} profiles in directions $\ell<$120\deg and 
$b>$0 are characteristically shaped by the LOS passing through, above, and then
plunging back through the flared outer disc at oblique angles \citep{fost03}. A 
good analytic figure for the warp is written as sinusoidal in azimuth $\phi$ 
and linear in $R$ ($m=$1), with a small quadratic ($m=$2) term:
\begin{equation}
z(R,\phi)=a_1(R-R_{max})\textrm{sin}\phi+a_2(R-R_{max})^2\textrm{sin}^2\phi
\label{warpeqn}
\end{equation}
\citet{bm98} give $a_1=$1/6, $a_2=$1/120 \& $R_{max}=$11~kpc, appropriate for 
a flat rotation curve with $R_0=$8.5~kpc \& 
$\Omega_0=$27.5~km~s$^{-1}$~kpc$^{-1}$. The observed $z>0$ displacement of the 
midplane is maximum towards $\ell=$80\deg, where the \ion{H}{i} layer extends
more than 20 degrees (3~kpc in $z$) from the equator. \citet{levi06a} fit 
$m=$0,1 \& 2 harmonic modes with the same equation, finding the $m=$1 mode 
begins inside the Solar circle and dominates to $R=$16~kpc, where the $m=$2 
mode begins to appear, introducing the asymmetry in the North and South halves 
of the warp. Observations that 3 out of 4 warped galaxies have no visible 
companions suggest that warps are not tidally excited by satellites. However, 
when dark matter is included in simulations with small companions ($\leq$2\% 
total mass), the {}``wake'' left by the orbiting companions in the DM halo 
induces vibration modes in the disc. In the MW, simulations by \citet{wein06} 
show a warp with 3 modes ($m=$0 1 \& 2) can be excited by the Magellanic clouds 
passing through the DM halo. 

The CGPS shows other essentially unstudied modes in the disc's topography with 
wavelengths of 5-20~kpc spectacularly well. These are collectively referred to 
as {}``scalloping'', and give the Outer disc a crinkly, corrugated appearance. 


\bibliography{foster_tyler}

\end{document}